\documentclass[12pt]{article}
\usepackage{amsmath, amsfonts}
\newcommand{\bee}{\begin{eqnarray*}}
\newcommand{\ene}{\end{eqnarray*}}
\newcommand{\beeq}{\begin{equation}}
\newcommand{\eneq}{\end{equation}}
\newtheorem{lem}{Lemma}[section]
\newcommand{\bel}{\begin{lem}}
\newcommand{\enl}{\end{lem}}
\newtheorem{defi}{Definition}[section]
\newcommand{\bef}{\begin{defi}}
\newcommand{\enf}{\end{defi}}
\newtheorem{exap}{Example}[section]
\newcommand{\beex}{\begin{exap}}
\newcommand{\enex}{\end{exap}} 
\newtheorem{theo}{Theorem}[section]
\newcommand{\beth}{\begin{theo}}
\newcommand{\enth}{\end{theo}}
\newtheorem{prop}{Proposition}[section]
\newcommand{\bep}{\begin{prop}}
\newcommand{\enp}{\end{prop}}
\newtheorem{cor}{Corollary}[section]
\newcommand{\bec}{\begin{cor}}
\newcommand{\enc}{\end{cor}}
\newtheorem{rem}{Remark}[section]
\newcommand{\ber}{\begin{rem}}
\newcommand{\enr}{\end{rem}}

\usepackage{amsmath}
\usepackage{graphicx}
\usepackage{latexsym}
\begin{document}
\title{ 
INFORMATION IN STOCK PRICES\\
AND\\
SOME  CONSEQUENCES: \\ A MODEL-FREE APPROACH
}
\author{Yannis  G. Yatracos\\
Cyprus U. of Technology
}
\maketitle
\date{}

{\it e-mail:} yannis.yatracos@cut.ac.cy

{\em Subject classifications:} Finance: Asset pricing. Games/group decisions: Bargaining. Statistics: Nonparametric.
\pagebreak
 
\begin{center}
{\bf Summary}
\end{center}

The price of a stock  will rarely follow the assumed model  and  a curious investor or a Regulatory Authority 
may wish to obtain a probability model the prices support.
 A  risk neutral probability
${\cal P}^*$  for the stock's  price at time $T$
is
determined  in closed form from 
 the prices before $T$ without 
assuming a price model.
Under mild conditions on the prices  the necessary and sufficient condition 
to obtain ${\cal P}^*$ 
is the coincidence at $T$  of the stock price ranges 
assumed  by the stock's trader and buyer.
This result 
clarifies the relation between market's  informational efficiency and the  arbitrage-free option pricing methodology. 
It also  shows  that in an incomplete market there are risk neutral probabilities
not supported by each stock and their use can be limited.
${\cal P}^*$-price $C$  for the stock's European  call option expiring at $T$  is 
obtained.  
Among other results it is shown for ``calm'' prices, like the log-normal,  that
{\em i)}
$C$
is the Black-Scholes-Merton price 
thus confirming  its 
validity for various stock prices,
{\em ii)} the  buyer's price carries an  exponentially increasing  volatility premium and its difference with $C$ provides a measure of the market risk premium.

\vspace{1in}


{\em Running Head:}
Information in stock prices

\pagebreak

\section{\bf Introduction}

\quad  In reality stock prices rarely follow the assumed model and research has been devoted in the past and  recently to discover
 the information these prices or the prices of stocks' derivatives  hide;
 see, e.g.,  Borovi\v{c}ka {\em et al.}
(2014), Ross (2015)  and references therein. 

An investor or a Regulatory Authority reluctant to follow 
the assumed price model  may 
wish to use
at  time $T$ risk neutral probability(-ies) ${\cal P}^*$  suggested by the stock prices
before $T.$  This is the central problem addressed in this work without 
price modeling assumptions. In a nutshell the  distribution of price returns is obtained as limit,  
under a 
sequence of probabilities determined by the prices,  and is 
modified
to become    risk neutral  ${\cal P}^*$
when  the stock's  trader and  buyer have the same belief/information
on the stock's potential values at $T.$
Thus,  in a complete market some of the available risk neutral probabilities are not supported by the prices of
 a particular stock and their use on this stock's derivatives are questionable.
The frequent quote
of the Black-Scholes-Merton ({\em B-S-M}) price (Black and Scholes,1973,  Merton 1973) is also  justified.

 The statistical consequences of these results are: I) the  L\'{e}vy triple characterizing ${\cal P}^*$ is given 
{\em in closed form}
via the stock prices before $T,$ thus it could be estimated non-parametrically, 
II) 
${\cal P}^*$'s  risk neutrality
is
equivalent to
 Le Cam's 
contiguity of  two sequences of probabilities 
 obtained from the stock prices before $T$ and III) the prices may support a mixture risk neutral probability with mixing
components determined from the prices' subsequences.
Due to I) this work complements 
 known results in the literature that guarantee only the existence of risk neutral probabilities.


More precisely,  mean-adjusted stock  prices and  Le Cam's theory of statistical experiments
 are  used
 to determine
one or more ${\cal P}^*$  at time $T$
via a constructive method.
The main assumption (${\cal A}3$) to obtain ${\cal P}^*$
 holds for  log-normal prices.
Each ${\cal P}^*$ is obtained from
a different  probability $Q$
with L\'{e}vy
 triple 
$[\mu_Q, \sigma^{2}_Q, L_Q].$ $Q$  is supported by the stock prices  before $T$ since 
 the triple's components 
are determined either 
by the sequence
of sums of  successive, adjusted  prices' jumps, or from one of   its subsequences.
${\cal P}^*$-price $C$  of the European call option  is obtained.
For  ``calm'' price, e.g. log-normal, with 
jumps not occuring often and  forcing  ${\cal P}^*$'s L\'{e}vy measure  to concentrate at zero,
$C$ 
is the {\em B-S-M} price,
thus confirming its  universal validity for various stock prices.
When the sequence of jumps' sums 
has subsequences converging weakly to different limits,  $C$ can be  obtained from  ${\cal P}^*$-mixtures  determined by the  cluster points of this sequence. 
The remaining available  risk neutral probabilities at $T$ correspond to stock prices with different jumps-variability
and should be used to price only  those stocks' derivatives.

 
 
Adopting the statistical experiments'  motivation from the 2-players' game, the European option's transaction is seen by the investor or the
Regulatory Authority as 
game between a trader and a buyer, both hypothetical. 
The buyer's price for the
option is shown to carry a volatility premium and  its difference from $C$ is a measure of the market's  risk premium.

A relation of the approach with the Kullback-Leibler relative entropy  is presented.
The connection of $Q$ and the obtained  ${\cal P}^*$ with 
 information from the stock prices  is revealed via
contiguity and is
related with the notion of  information used in the Market Manipulation literature; see, e.g., Cherian and Jarrow (1995).
$Q$  is defined via  sequences of beliefs-probabilities      
$\{P_{tr,n}\}$ and $\{P_{bu,n}\},$ respectively,
of the trader and of the buyer,
$n$ determines the number of stock  prices providing information before $T$ and $P_{tr,n}$ is equivalent to $P_{bu,n}$  for every $n \ge1.$
$P^*$ is risk neutral, if and only if, 
$\{P_{tr,n}\}$ and $\{P_{bu,n}\}$ are contiguous, implying that
with  information  from countably infinite stock prices  the  beliefs-probabilities of these agents for $S_T$ 
remain  equivalent.
Thus, neither the trader nor the buyer
 have private information on $S_T$'s range of values.
This result complements those in Jarrow (2013) which demonstrate the intimate relationship between an informationally efficient market and option pricing theory. It also confirms  the Third Fundamental Theorem of Asset Pricing 
(Jarrow and Larsson, 2012, Jarrow, 2012, Jarrow, 2013, Corollary (Market Efficiency) p. 91) 
by providing sufficient conditions under which the family of stock price returns is an information set that makes 
the market efficient.

The foregoing results appear in sections 4 and  5. The original theoretical contributions consist of
{\em i)} the new 2-step method proposed in section 2 to determine ${\cal P}^*$
via  $Q$   which
motivates in section 4 the
embedding in
the statistical experiments
framework
 (see (\ref{eq:statexp101})), and {\em ii)}  Propositions 4.1-4.4,   Lemma 6.1 in  Appendix 1 and Proposition 7.1 in Appendix 2.
These are used in the applications section 5 to present new, quantitative results. 
The tools
 in section 3  include mean-adjusted  price $p_t(=S_t/E_{\cal P}S_t)$
which is density 
on the probability space $(\Omega, {\cal F}, {\cal P}); \ t$  denotes time, $ t_0 \le  t \le T.$ 
 Beliefs-probabilities $P_{tr,n}$ and $P_{bu,n}$
are 
$p_t$-products
 at various trading times 
$t$  in  the  interval $[t_0,T].$

For the reader who feels uncomfortable 
because
$\{P_{tr,n}\}$ and $\{P_{bu,n}\}$ 
are product of prices-densities
and indicate  independence of price returns, it should be reminded that  Fama's weak Efficient Market Hypothesis  implies 
either independence or slight dependence of the stock price returns
(Fama, 1965,  p. 90, 1970, pp. 386, 414). Modeling ``slight" dependence with  weak dependence  is acceptable in  Finance (Duffie, 2010, personal communication). Thus, the limiting laws  obtained 
under $\{P_{tr,n}\}$ and $\{P_{bu,n}\}$  remain valid under weak dependence and the 
obtained results hold.

Janssen and Tietje (2013) use statistical experiments to discuss ``the connection between mathematical
finance and statistical modelling''
 for $d$-dimensional price processes.
Some of the differences in their work are: {\em a)} The  price process  is not standardized and ${\cal P}^*$  is assumed to exist.
 {\em b)} Convergence of the likelihood ratios to a normal experiment is obtained under the contiguity assumption. 
{\em c)} The relation between ${\cal P}^*,$  market's informational efficiency and  contiguity is not revealed.
{\em d)} There are no results explaining the behaviors of the trader and of the buyer.
{\em e)} There is no proof of the universal validity of  {\em B-S-M} formula  without
price modeling assumptions.


The theory of statistical experiments 
used 
is in Le Cam (1969, Chapters 1 and 2, 1986, Chapters 10 and 16),
Le Cam and Yang (1990, Chapters 1-4, 2000, Chapters 1-5)
 and in Roussas (1972, Chapter 1).
 Theory of option pricing can be found, e.g., in
Musiela and Rutkowski (1997). 
A concise and very informative presentation of L\'{e}vy processes theory can be found in Kyprianou (2006).
Proofs and auxiliary results are in Appendices  1 and 2. 



\section{The approach  to obtain ${\cal P}^*$ via $Q$}

\quad  
${\cal P}^*$ to be used at $T$  is equivalent to the physical probability ${\cal P}$ and satisfies the equation: 
\begin{equation}
\label{eq:weneed}
E_{{\cal P} ^*}(\frac{S_T}{S_t}|{\cal F}_t)
=e^{r(T-t)}; 
\end{equation}
stock prices $\{S_t, t>0\}$ are defined on the probability space $(\Omega, {\cal F}, {\cal P}),$  $\{ {\cal F}_t\}$ is the natural filtration and  $t$ denotes time.
We consider a re-expression of (\ref{eq:weneed}),
\begin{equation}
\label{eq:weneedreexp}
E_{{\cal P} ^*}(\frac{S_T}{S_t}|{\cal F}_t)=E_{Q^*}[\exp\{V_{T-t}+ \ln a_{[t,T]}\}|{\cal F}_t]=e^{r(T-t)},
\end{equation}
\begin{equation}
\label{eq:Tools}
V_{T-t}=\ln \frac{S_T/ES_T}{S_t/ES_t}, 
\end{equation}
\begin{equation} 
\label{eq:a}
 a_{[t,T]}=\frac{ES_T}{ES_t};
\end{equation}
$ES_t$ denotes $E_{\cal P}S_t$ for every $t$-value. $Q^*$ is the cumulative distribution function (c.d.f.) of $V_{T-t}$ under ${\cal P}^*$  and will be obtained {\em in two steps}.
Unless needed, domains of integration are omitted since they are determined by the  c.d.fs./probabilities.\\

{\bf Step 1:}
Determine  $Q$ for
$V_{T-t}$
 under which 
$\{S_t/ES_t\}$
is a martingale, i.e.
\begin{equation}
\label{eq:Pln}
E_Q(\exp \{
V_{T-t}\}|{\cal F}_t)=\int e^{v} dQ(v|{\cal F}_t)=1.
\end{equation}
 There is no involvement of the interest (i.e. of  $r$)  in Step 1.\\

{\bf Step 2:}  $Q^*$ is  the translated  $Q,$
\begin{equation}
\label{eq:Q*}
Q^*(v|{\cal F}_t)=Q[v+\ln a_{[t,T]}-r(T-t)|{\cal F}_t].
\end{equation}
\ber \label{r:step1impor} 
$Q$  is the key element that allows to obtain ${\cal P}^*$ without making model assumptions, 
and  reveals ${\cal P}^*$'s  relation with the flow of information.
When $Q$ is determined,  (\ref{eq:weneedreexp}) holds  under $Q^*:$ 
$$
E_{Q^*}[e^{V_{T-t}+\ln a_{[t,T]}}|{\cal F}_t]
=\int e^{v+\ln a_{[t,T]}} dQ[v+\ln  a_{[t,T]} -r(T-t)|{\cal F}_t]
$$
$$= \int e^{w+r(T-t)} dQ(w|{\cal F}_t)=e^{r(T-t)}.$$
\enr




 For Geometric Brownian motion the 2-step approach allows to obtain  $Q.$

\beex
\label{ex:lognormBSM}
  Let $S_t$ be a geometric Brownian motion,
\begin{equation}
S_t=s_0exp\{(\mu-\frac{\sigma^2}{2})t+\sigma B_t\}
\label{eq:gbm}
\end{equation}
with $B_t$ standard Brownian motion, $t>0$ and $s_0$ the price at
$t=0.$
Since $ES_t=s_0exp\{\mu t\}$
$$
V_{T-t}
=-\frac{\sigma^2}{2}(T-t)+ \sigma (B_{T}-B_{t}), \ t<T,
$$
and  (\ref{eq:Pln}) holds
under ${\cal P},$ i.e.,  $Q$   coincides with  ${\cal P},$ and (\ref{eq:weneedreexp})  holds under $Q^*.$  

 
\enex

In Example \ref{ex:lognormBSM}, $Q, \ {\cal P}^*$ and $Q^*$ are easily obtained  because the distribution of 
$V_{T-t}$
 is normal.  Can one similarly obtain $Q$ (and therefore $Q^*, P^*$)  in other situations? 
Without stock price modeling assumptions  
the  stock  prices 
before $T$ provide information 
to  the investor and/or the Regulatory Authority
and  can determine  
 $Q$ 
via a sequence $Q_n;$  
$n$ increases with the flow of information, i.e., the number of stock prices  in $(t,T).$ 
This
 is supported by  Cox, Ross and Rubinstein (1979) where, for the binomial  price model,  {\em B-S-M} price  is obtained as limiting price. 
Also, by the terms
$(\stackrel{+}{-}\ .5 \sigma^2)$ in the standard normal c.d.fs. of 
{\em B-S-M} price, 
indicating these  c.d.fs. 
are  limits  of expected values under contiguous sequences of probabilities. 


 When there are  ``many'' transactions in  $[t,T]$  an embedding in Le Cam's statistical experiments, where contiguity was introduced, 
allows to determine $Q.$ 
\ber \label{r:stochint1} When $r$ is not fixed in $[t,T],$ let $B_{[t,T]}=e^{\int_{t}^Tr_s ds};$  $r_s$ determines  the interest  at time $s,$ as $r$ determined interest $i.$ Then in (\ref{eq:weneed}) and  (\ref{eq:weneedreexp}), $B_{[t_0,T]}$ is replacing  $e^{r(T-t_0)}$ and in  (\ref{eq:Q*}),
 $\ln B_{[t_0,T]}$ is  replacing $r(T-t_0).$ All subsequent results herein still  hold with these changes.
 \enr
 
\section{Binary Statistical Experiments, Contiguity 
and  Market's  Informational  Efficiency }


\quad A {\em binary statistical experiment} ${\cal E}$ consists of probabilities
$\{Q_1, Q_2\}$
on $(\tilde \Omega, \tilde {\cal F})$ (Blackwell, 1951).
Le Cam (see, e.g., 1986) 
 introduced 
a distance $\Delta$ 
between experiments 
and
proved that {\em $\Delta$-convergence}  of binary experiments
 ${\cal E}_n=\{Q_{1,n}, Q_{2,n}\}, \ n\ge 1,$
to ${\cal E}$
is equivalent to
weak convergence of likelihood ratios
$\frac{dQ_{2,n}}{dQ_{1,n}}$ under $Q_{1,_n}$ (resp. $Q_{2,n}$)  to the distribution
of $\frac{dQ_2}{dQ_1}$ under $Q_1$ (resp. $Q_2$).

From several equivalent definitions of contiguity we present one that will 
 reveal 
the relation between  ${\cal P}^*$ 
and the stock prices' information.
\bef 
\label{f:defcon} 
 (see, e.g. Le Cam, 1986, p. 87, Definition 5) Let  ${\cal E}_n=\{Q_{1,n}, Q_{2,n}\}, \ n\ge 1,$
be a sequence of statistical experiments. Then, the sequence $\{Q_{1,n}\}$ is contiguous to the sequence$\{Q_{2,n}\}$
if in all cluster points ${\cal E}=\{Q_1, Q_2\}$ of the sequence  $\{{\cal E}_n\}$  (for $\Delta$-convergence), $Q_1$ is dominated by $Q_2.$

If, in addition, $\{Q_{2,n}\}$ is contiguous to sequence$\{Q_{1,n}\}$ then the sequences $\{Q_{1,n}\}, \ \{Q_{2,n}\}$ are contiguous.
\enf
\quad From Definition \ref{f:defcon},  for contiguous sequences of probabilities $\{Q_{1,n}\}, \{Q_{2,n}\}$  forming
a sequence of binary experiments ${\cal E}_n,$  each 
 cluster point experiment of $\{{\cal E}_n\}$ consists of mutually absolutely continuous probabilities.

The definition of an informationally efficient  market  has been recently  formalized.
\bef 
\label{d:mefjl} (Jarrow and Larsson, 2012, Jarrow, 2013, p. 89) A market is efficient with respect to an information set 
${\cal F}^*$ for which the price process is measurable,  if there exists an equilibrium economy (with supply equal to demand) that supports the market's price process 
and where the price process reflects the information set ${\cal F}^*.$
\enf
\quad An important result follows relating market's  informational efficiency with  risk neutral probabilities.

\bep
\label{eq:mefprop} (Jarrow, 2013, p. 90)
The market is efficient
with respect to an information set ${\cal F}^*$  if and only if there exist
risk-neutral probabilities ${\cal P}^*$  such that  $S_t/B_t$ is a martingale.
\enp
\quad Herein  the information set ${\cal F}^*$ of Proposition \ref{eq:mefprop}    is determined by the successive stock  price returns which provide ${\cal P}^*.$



\section{Modeling 
$V_{T-t_0}$}

\subsection{The Embedding}

\quad We  introduce binary statistical experiments determined by the stock prices.

Consider 
the process of
prices-densities
\begin{equation}
\{p_t
=\frac{S_t}{ES_t}, \ t \in [0,T]\}.
\label{eq:pricedensities}
\end{equation}
Embed
 the stock prices in $[t_0,T]$  in the
statistical 
experiments' framework
by re-expressing 
$V_{T-t_0}$ in (\ref{eq:Tools})
 using  the intermediate prices-densities,
\begin{equation}
V_{T-t_0}= \ln \frac{S_T/ES_T}{S_{t_0/ES_{t_0}}}
=\ln \frac{\frac{S_T}{ES_T} \ldots \frac{S_{t_1^n}}
{ES_{t_1^n}}}
{\frac{S_{t_0}}{ES_{t_0}} \ldots \frac{S_{t^n_{k_n-1}}}{ES_{t^n_{k_n-1}}}}
=\ln \frac{\Pi_{j=1}^{k_{n}} p_{t_j^{n}} } {\Pi_{j=1}^{k_{n}} p_{t_{j-1}^n} }
=\ln \Pi_{j=1}^{k_{n}} \frac{  p_{t_j^{n}} }{ p_{t_{j-1}^n} }
=\Lambda_{k_n}.
\label{eq:embedding}
\end{equation}

The products 
$\Pi_{j=1}^{k_{n}}
p_{t_{j-1}^n}$ and $\Pi_{j=1}^{k_{n}}
p_{t_j^{n}}$
 determine, respectively, 
beliefs-probabilities   $P_{tr,n}$ and $ P_{bu,n}$
 in $(\Omega^{k_n}, {\cal F}^{k_n})$
and the statistical experiment
\begin{equation}
\label{eq:statexp101} 
 {\cal E}_{n}=\{P_{tr,n}
\mbox{   } P_{bu,n}
 \};
\end{equation} 
$t_0^n=t_0$ and $t_{k_n}^n=T$ for each $n.$ 
$P_{tr, n}$ and $P_{bu, n}$ are both unknown but when $n$ and so $k_n$ increase to infinity, with mild assumptions,
the theory provides the asymptotic distributions of $\ln \frac{dP_{bu,n}}
{dP_{tr,n}}$ under both $P_{tr,n}$ and $P_{bu,n}.$

 It  is shown
that when the trading times are dense in $[t_0,T]$  the  distributions of 
$\log \frac{p_T}{p_{t_0}}$ under  $P_{tr,n}$ and $P_{bu,n}$ are infinitely divisible, normal in particular for calm stock, and $Q$ is obtained (Propositions \ref{t:LLCmain}, \ref{p:QQ*}  and  Corollary \ref{t:asydicalm}). 
 When $t_0$ is the present and $S_{t_0}=s_{t_0},$ $\frac{p_T}{p_{t_0}}$ becomes  $p_T$  and
${\cal P}^*$  obtained via $Q$  is  $S_T$'s probability  (Proposition \ref{p:STpstar}).
 
\subsection{The Assumptions}

\quad  Let $Y_{n,j}$ be  a  fluctuation measure of $\frac{p_{t_j^n}}{ p_{t^n_{j-1}}}$ from unity,
 \begin{equation}
Y_{n,j}=\sqrt { \frac {p_{t_j^n}}{p_{t^n_{j-1}}}} -1,  \ j=1,\ldots, k_n.
 \label{eq:yu}
\end{equation}
Assume 

${({\cal A}1)}$  $S_t>0$ and $ES_t<\infty$ for
every $t \in [t_0,T],$

${({\cal A}2)}$  a countable number of stock's transactions 
in any open interval
of $[t_0,T],$

${({\cal A}3)}$ for 
the prices-densities $p_{t^n_0}, \ p_{t_1^n}, \ldots, p_{t^n_{k_n}}$
and  mesh size $\delta_n=\sup \{t_j^n-t^n_{j-1}; j=1,\ldots, k_n\}, \
k_n=k_n(\delta_n), t_0^n=t_0, t_{k_n}^n=T, \ d P_{t_j^n}= p_{t_j^n}d{\cal P}, \ j=1,\ldots, k_n,$
\[
(i) 
\lim_{\delta_n\rightarrow 0} \sup    \{ E_{P_{t^n_{j-1}}} Y_{n,j}^2,
\ j=1,\ldots, k_n\}=0,
\]
\[ 
(ii)  \mbox{ there is positive $b:$}
\hspace{2ex}\sup_n\sum_{j=1}^{k_n} E_{P_{t^n_{j-1}}}  Y_{n,j}^2 \le b<\infty.
\]
${({\cal A}4)}$  Under ${\cal P},$
$\lim_{n \rightarrow \infty} S_{t_1^n}=s_{t_0}$ in probability,
$\lim_{n \rightarrow \infty} ES_{t_1^n}=s_{t_0}.$ 

Conditions are provided in Appendix 2 
for ${\cal A} 3 $ to hold. In Proposition \ref{p:A3calmgbm} it is shown that ${\cal A} 3 $  holds for Geometric Brownian motion. Condition ${\cal A} 3 (ii) $ is weaker than  boundedness  of the expected quadratic variation of $\sqrt{p_t}, \ t \in [t_0,T],$ 
under ${\cal P}.$

Assumption ${{\cal A}1}$ allows
the passage from stock prices to prices-densities and $-1<Y_{n,j}<\infty,  j=1,\ldots, k_n.$
Assumptions ${{\cal A}1}$ and ${{\cal A}2}$ provide sequences of 
probabilities with $n$-th terms,
$P_{tr,n}$  and $P_{bu,n},$ 
mutually absolutely continuous.
Assumption ${\cal A}3 (i)$ indicates that 
the contribution of the ratio 
$\frac{p_{t_j^n}}{p_{t_{j-1}^n}}$
does not affect the distribution of 
$V_{T-t_0}, \ j=1,\ldots,k_n.$ 
Assumption ${\cal A}3 (ii)$ 
 implies 
that
$V_{T-t_0}$'s variance is finite. It is a condition involving the square roots of the stock prices and
does not guarantee their square integrability which would allow showing, via the martingale convergence theorem, that
the discounted stock prices are a L\'{e}vy process. Assumption ${\cal A}3 (ii)$ can be relaxed 
following Lo\`{e}ve (1977) leading to a different representation of the moment generating or the characteristic function in
Proposition \ref{t:LLCmain}.
 Assumption ${\cal A}4$ allows, along with $Q,$  to change the conditional expectation in 
(\ref{eq:weneed}) to expected value. Most important, ${\cal A}4$  allows to show that when
$t_0$ is the present and  $S_{t_0}=s_{t_0}$ then
$V_{T-t_0}=\ln (S_T/ES_T)$  and $Q$ determines $P^*$  for $S_T.$ 
 

\subsection{\bf The $Q$-distribution(s)  of
$V_{T-t_0}$
}

\quad Assumption ${\cal A}3$ implies also that under $\{P_{tr,n}\}$ and $\{P_{bu,n}\}$ the 
sequences of 
distributions of $\sum_{j=1}^{k_n}Y_{n,j}$ 
are relatively compact.
 Thus, we can choose a subsequence $\{k_{n'}\},$  
for which  $\sum_{j=1}^{k_{n'}}Y_{n',j}$ 
converges weakly, respectively, under $P_{tr,n'}$ and $P_{bu,n'},$ to infinitely divisible distributions.
 Without loss of generality we  use $\{n\}$ and $\{k_n\}$
instead of $\{n'\}$ and $\{k_{n'}\}.$ If there are two or more subsequences with different weak  limits,
the stock prices support more than one risk neutral  probabilities and ${\cal P}^*$ can be modeled as mixture of infinitely divisible distributions. 

 $V_{T-t_0}$  is approximated in probability by a linear function of  $\sum_{j=1}^{k_n}Y_{n,j}$
and has 
under $P_{tr,n}$ as limit the  infinitely divisible distribution $Q_0.$
 A translation of  $Q_0$ is usually needed  to obtain $Q$ satisfying (\ref{eq:Pln}).

The next proposition for the binary experiment (\ref{eq:statexp101}) follows using a result  from Le Cam (1986, Proposition 2, p. 462) that holds for two product probabilities forming a likelihood ratio.

\bep \label{t:LLCmain}
Assume that ${\cal A}1-{\cal A}3$ hold and that  $\sum_{j=1}^{k_n}Y_{n,j}$   has under $P_{tr,n}$ a weak limit with
L\'{e}vy  triple  $[\mu, \sigma^2, L_{tr}].$ 
Under  $P_{tr,n}, \  \Lambda_{k_n}=\ln \Pi_{j=1}^{k_n}\frac{p_{t_{j}^n}}{p_{t_{j-1}^n}}$ 
converges in distribution to $\Lambda$
with c.d.f.   $Q_0$
and for every $s \in (0,1)$ its moment generating function,
\begin{equation}
\psi_{Q_0}(s)=\ln E_{Q_0}e^{s\Lambda}
=\mu_{[t_0,T]}s+\frac{\sigma^2_{[t_0,T]}}{2}s^2
+\int_{[-1,0)\cup (0,\infty)}
[(1+y)^{2s}-1-2sy]L_{tr}(dy);
\label{eq:mgflambda}
\end{equation}
$$\mu_{[t_0,T]}=2\mu-\sigma^2<0, \hspace{5ex}  \sigma^2_{[t_0,T]}=4\sigma^2,$$
 and  $\mu, \sigma^2, L_{tr}$
are all  determined in Remark \ref{r:lambda}. \\
\enp

\ber \label{r:lambda}
{Under ${\cal A}1,$ in (\ref{eq:mgflambda}) $L_{tr}(-1)=0.$  
The model parameters 
$$
\mu_{[t_0,T]}=2\mu-\sigma^2=(2\mu_1-\sigma_1^2)(T-t_0), \hspace{4ex} \sigma^2_{[t_0,T]}=4\sigma^2=4\sigma_1^2(T-t_0);
$$
 $\mu_1$ and $\sigma_1^2$ are determined from an interval of length unity,
\begin{equation}
\label{eq:unitparam}
\mu_1=\lim_{n \rightarrow \infty}\sum_{j=1}^{k_n}E_{P_{t_{j-1}^n}}Y_{n,j}<0, \ \hspace{5ex} 
\sigma_1^2=
\lim_{\tau \rightarrow 0} \lim_{n \rightarrow \infty} \sum_{j=1}^{k_n}
E_{P_{t_{j-1}^n}}Y_{n,j}^2I(|Y_{n,j}|\le \tau).
\end{equation}
The L\'{e}vy measure in an interval of length unity is\\ 
\begin{equation}
L_{tr}(y)
=\lim_{n \rightarrow \infty}\sum_{j=1}^{k_n}
E_{P_{t_{j-1}^n}}Y_{n,j}^2I(Y_{n,j}\le y).
\label{eq:unitLevy}
\end{equation}
 
}
\enr

From  L\'{e}vy-Khintchine theorem (see , e.g., Kyprianou,  2006,
Theorem 2.1, p. 35),  $V_{T-t}$ can be seen as L\'{e}vy process 
 thus the conditional expected value in (\ref{eq:weneedreexp}) is an expected value; see also  Lemma \ref{p: nocond}
 in  the Appendix.

The next proposition provides $Q$ and $Q^*$ 
(via $Q_0$) and the necessary and sufficient condition (\ref{eq:neutraliff}) 
to obtain $Q$ which is equivalent to contiguity of the sequences $\{P_{tr,n}\}$ and $\{P_{bu,n}\}.$
 $I$ denotes indicator function.

\bep
\label{p:QQ*}
 a) $Q_0$ in Proposition \ref{t:LLCmain} satisfies  (\ref{eq:Pln}), if and only if,
\begin{equation}
\label{eq:neutraliff}
\mu_{[t_0,T]}+.5 \sigma_{[t_0,T]}^2+E_{L_{tr}}Y^2 I(Y\neq 0)=2\mu+\sigma^2+E_{L_{tr}}Y^2 I(Y\neq 0)=0.
\end{equation}
b) $Q$ satisfying  (\ref{eq:Pln}) has  L\'{e}vy 
triple  $[-.5 \sigma_{[t_0,T]}^2-  E_{L_{tr}}Y^2 I(Y\neq 0), 
\sigma_{[t_0,T]}^2, L_{tr}],$
\begin{equation}
\label{eq:Q}
 Q(v)=
\int \Phi(\frac{v+.5\sigma_{[t_0,T]}^2+ E_{L_{tr}}Y^2I(Y\neq 0) -y}{\sigma_{[t_0,T]}})L_{tr, Pois}(dy);
\end{equation}
$\Phi$ denotes the c.d.f. of a standard normal random variable, $L_{tr, Pois}$ is the probability of the Poissonian component. The triple's parameters can be estimated using Remark \ref{r:lambda}.\\
c) 
Let  $Q_{bu}$ be the limit distribution of $\Lambda_{k_n}$ under $P_{bu,n}.$ 
$\{P_{tr,n}\}$ and $\{P_{bu,n}\}$ are
contiguous (and therefore $Q_0, Q_{bu}$ in Proposition \ref{t:LLCmain}  are mutually absolutely continuous),   if and only if,
(\ref{eq:neutraliff}) holds.
Then, $Q_{bu}$  has   L\'{e}vy 
triple  $[.5 \sigma_{[t_0,T]}^2+ E_{L_{tr}}Y^2I(Y\neq 0),\sigma_{[t_0,T]}^2, L_{bu}],$
\begin{equation}
\label{eq:QB1}
Q_{bu}(v)=
\int \Phi(\frac{v-.5\sigma_{[t_0,T]}^2- E_{L_{tr}}Y^2I(Y \neq 0)-y}{\sigma_{[t_0,T]}})L_{bu,Pois} (dy);
\end{equation}
$L_{bu,Pois}$ is the probability of the Poissonian component.\\
d) For $Q$ in b), 
 \begin{equation}
\label{eq:QB2}
dQ_{bu}(v)=e^v dQ,
\end{equation}
\begin{equation}
\label{eq:ICequ}
\int_{\{v:v>x\}} e^v dQ(v)=1-Q_{bu}(x), \ \forall x \in R.
\end{equation}
e)  
$Q^*$ is obtained using 
$Q$ in b) 
and 
(\ref{eq:Q*}).
\enp


\ber
From (\ref{eq:neutraliff}), $Q_0$ is risk neutral when the drift $\mu$ equals $-.5\sigma^2 - .5 E_{L_{tr}}Y^2I(Y\neq 0).$ 
There are several possible $\tilde \sigma$ values and L\'{e}vy measures $\tilde L_{tr}$ for which  (\ref{eq:neutraliff}) holds,
but those supported by the stock prices (at unit time length) are  given by (\ref{eq:unitparam}) and  
(\ref{eq:unitLevy}).
The remaining $\tilde L_{tr}$ and $\tilde \sigma$  correpond to stock prices with different jumps-variability.
\enr

The result allowing to obtain ${\cal P}^*$ for $S_T$  via $Q$ follows.
 
\bep
\label{p:STpstar} Assume that ${\cal A}1-{\cal A}4$ hold for stock prices $\{S_t, t_0<t\le T\}.$ When $t_0$ is the present, $S_{t_0}=s_{t_0},$  Proposition \ref{p:QQ*}  holds for $V_{T-t_0}=\ln(S_T/ES_T)$ and $Q, Q^*, {\cal P}^*$ are obtained.
\enp




\subsection{\bf  The $Q$-distribution(s) of 
$V_{T-t_0}$ for Calm Stock}


\quad Calm stock
 has 
prices-densities $p_{t+\delta}, \ p_t$ (see (\ref{eq:pricedensities}))
that do not differ often much 
 with respect 
to ${\cal P}$ 
for small $\delta$-values.

\bef
{\em Let $t_1^n<\ldots< t^n_{k_n-1}$ be a partition  of $(t_0=t_0^n, T=t^n_{k_n}),$
with mesh size $\delta_n=\sup \{t^n_j-t^n_{j-1}, \ j=1,\ldots, k_n\}.$ 
{\em Stock $\{S_t\}$ is 
calm}  in $[t_0,T]$ if for any $\epsilon (>0)$ and any partition
 \begin{equation}
\lim_{\delta_n \rightarrow 0 }  \sum_{j=1}^{k_n}E_{P_{t^n_{j-1}}}
(\sqrt{\frac{p_{t_j^n}}{p_{t^n_{j-1}}}}-1)^2
I(|\sqrt{\frac{p_{t_j^n}}{p_{t^n_{j-1}}}}-1|>\epsilon)=\lim_{\delta_n \rightarrow 0 }  \sum_{j=1}^{k_n}E_{P_{t^n_{j-1}}}
Y_{n,j}^2
I(|Y_{n,j}|>\epsilon)
=0;
\label{eq:ecalmexp}
\end{equation}
$I$ is the indicator function.
}
\enf

When ${\cal A} 3 (i)$ holds, the calm stock condition  for the
 random variables 
$Y_{n,j}, \ j=1,\ldots,k_n$ 
guarantees that $\sum_{j=1}^{k_n}Y_{n,j}$ has
asymptotically  normal distribution (LeCam, 1986, p. 470) and
the same holds for
$$\Lambda_{k_n}=2\sum_{j=1}^{k_n}  \ln (1+Y_{n,j}),$$
i.e., for $V_{T-t_0}.$


\bec \label{t:asydicalm}
When ${\bf {\cal A}}1, {\bf {\cal A}}2, \mbox{ and }{\bf {\cal A}}3 (i)$
hold for a calm stock in  $[0,T],$  for every convergent subsequence $\sum_{j=1}^{k_{n'}}Y_{n',j}$
there is $\sigma_{[t_0,T]}>0$
such that\\
 \begin{equation}
(i)
\hspace{2ex}
Q(v)=
\Phi(\frac{v+\frac{\sigma^2_{[t_0,T]}}{2}} {\sigma_{[t_0,T]}}),
\label{eq:traderprob}
\end{equation}
\begin{equation}
(ii)\hspace{2ex} Q_{bu}(v)=
\Phi(\frac{v-\frac{ \sigma^2_{[t_0,T]}}{2}} {\sigma_{[t_0,T]}}).
\label{eq:buyerprob}
\end{equation}

When, in addition, 
$\sum_{j=1}^{k_n}E_{P_{t_{j-1}^{n}}}Y_{n,j}^2$
 has a limit as $n$ increases to infinity then $Q$ and so $P^*$ are uniquely determined.
\enc


Proposition \ref{p:A3calmgbm} shows that assumption ${\cal A}3$ holds for the Geometric Brownian motion model 
which is calm. 

\bep
\label{p:A3calmgbm}
For the price-densities of the Geometric Brownian motion (\ref{eq:gbm}):\\
{a)} assumption ${\cal A}3$ holds  and
\begin{equation}
\label{eq:sumexpconv}
\lim_{\delta_n \rightarrow 0 }  \sum_{j=1}^{k_n}E_{P_{t^n_{j-1}}}
Y_{n,j}^2=\lim_{\delta_n\rightarrow 0} \sum_{j=1}^{k_n}[\int (\sqrt{p_{t_j}}-\sqrt{p_{t_{j-1}}})^2d{\cal P}
=.25 \sigma^2 (T-t_0),
\end{equation}
 {b)} the calm stock condition (\ref{eq:ecalmexp}) also holds.
 \enp



\section{ \bf Applications and Consequences}

\subsection{Option pricing of a European call}

For the hypothetical trader and the buyer make the usual assumption:

 ${({\cal A}5)}$ The market 
consists of the stock $S$ and a risk-less bond
that appreciates at fixed rate $r$ and there are no dividends or
transaction costs. The option is European. The buyer prefers to 
pay less than more. 
 

${\cal P}^*$-price of a European call is obtained from a weakly convergent subsequence of  $\sum_{j=1}^{k_n}Y_{n,j}.$ 
When  $\sum_{j=1}^{k_n}Y_{n,j}$ has several  cluster points,
the fair price is a weighted sum
of the corresponding ${\cal P}^*$-prices. 

\bep \label{p:priLevy} Assume
${\cal A}1- {\cal A}5$  hold.
 The ${\cal P}^*$-price $C$  of the European call option at $t_0,$ with strike price $X$ at expiration $T,$ 
  is
 \begin{equation}
\label{eq:Eprice}
C=s_{t_0}R_{bu}-X e^{-r(T-t_0)} R_{tr}, 
\end{equation}
\begin{equation}
\label{eq:buyerprobEnd1} 
R_{bu}=1-Q_{bu}[\ln(X/s_{t-0})-r(T-t_0)]
\end{equation}
$$
=\int \Phi(\frac{\ln(s_{t_0}/X)+r(T-t_0)+.5\sigma_{[t_0,T]}^2+ E_{L_{tr}}Y^2I(Y \neq 0)+y}{\sigma_{[t_0,T]}})L_{bu,Pois}(dy)
$$
$$
R_{tr}=\int \Phi(\frac{ \ln (s_{t_0}/X)+r(T-t_0)-.5\sigma_{[t_0,T]}^2- E_{L_{tr}}Y^2I(Y \neq 0)+y}
{ \sigma_{[t_0,T]} })
L_{tr,Pois}(dy).
$$
\enp

For calm stock, B-S-M price is obtained without model assumptions
thus justifying its universality and frequent quote.

\bec \label{c:EpriceCalm} For calm stock, under the assumptions ${\cal A}1, {\cal A}2, {\cal A}3 (ii), {\cal A}4, {\cal A}5,$   the 
coefficients  $R_{bu}$ and $R_{tr}$ in (\ref{eq:Eprice}) are
$$ R_{bu}=\Phi(\frac{\ln(s_{t_0}/X)+r(T-t_0)+.5\sigma_{[t_0,T]}^2}{\sigma_{[t_0,T]}}), $$
$$ R_{tr}=\Phi(\frac{\ln(s_{t_0}/X)+r(T-t_0)-.5\sigma_{[t_0,T]}^2}{\sigma_{[t_0,T]}}).$$
\enc

 \subsection{\bf Binary Statistical Experiments, Information, Market's Efficiency  and ${\cal P}^*$}

\quad We  relate the approach in this work 
 with
 notions of information. It is assumed that the investor and/or the Regulatory Authority do not know the stock's price model and 
the prices' successive returns provide  information for $P^*$ and  the beliefs of the trader and the buyer.
\bef
\label{d:KL} (see, e.g., Cover and Thomas, 2005, p. 19) The relative entropy, or Kullback-Leibler distance, between two densities
$f$ and $g$ is
\begin{equation}
\label{eq:KL}
D(f||g)=-E_f\ln \frac{g(X)}{f(X)}.
\end{equation}
\enf
\quad Herein 
we use  the distribution of  $\ln \frac{g(X)}{f(X)}$ under $f$  rather than  its  expected value  (\ref{eq:KL}).
In our notation $f, \ g$ are, respectively, either $P_{tr,n},P_{bu,n}$ or   $Q,Q_{bu}.$

In each of the binary experiments ${\cal E}_n=\{P_{tr,n}, P_{bu,n}\}$ and  $\{Q,Q_{bu}\},$  the beliefs-probabilities for $S_T$'s distribution are, respectively, those of the trader and the buyer.  $P^*$ is determined via $Q$ that satisfies (\ref{eq:Pln})  if and only if  (\ref{eq:neutraliff}) holds. The latter equation is equivalent to contiguity of the sequences $\{P_{tr,n}\}$  and $\{P_{bu,n}\};$ 
see  Proposition \ref {p:QQ*} c).
Thus, the obtained ${\cal P}^*$ is risk neutral, if and only if, $\{P_{tr,n}\}$  and $\{P_{bu,n}\}$ are contiguous.
 To see what this means 
for the trader and the buyer observe that  for each $n,$ 
$P_{tr,n}$  and $P_{bu,n}$ are mutually absolutely continuous  
 and  are based on information from  $k_n$ stock prices
before
$T$  for determining 
$S_T$'s distribution. 
Therefore, the corresponding  induced probabilities for $\Lambda_{k_n},$$P_{tr,n}\circ \Lambda_{k_n}^{-1}$   
and $P_{bu,n}\circ  \Lambda_{k_n}^{-1},$  are also mutually absolutely continuous; 
see (\ref{eq:embedding}) for $\Lambda_{k_n}.$
Proposition \ref{p:QQ*} shows that with infinite amount of information,  i.e., when $n$ and $k_n$ increase to infinity, the (limit) beliefs-distributions $Q$ and $Q_{bu}$ for  $S_T$ are also mutually absolutely continuous.
Thus, neither the trader nor the buyer have private information on $S_T$'s values any time in $(t_0,T).$
Proposition  \ref{eq:mefprop} implies that the market is efficient with respect to the information set
${\cal F}^*$ determined by the stock price returns in $[t_0,T)$ which provide ${\cal P}^*.$ Thus the results are  a compagnon of the Third Fundamental Theorem of Asset Pricing
 (Jarrow, 2012) which characterizes the conditions under which an equivalent martingale probability measure exists in the economy.
Recall that the above hold for each convergent subsequence $\{{\cal E}_{n'}\}$ (see (\ref{eq:statexp101})).

Mutual absolute continuity of traders' beliefs-probabilities
 is used in Financial Economics, e.g.,  the area of Market Manipulation.  Cherian and Jarrow (1995, p. 616, Assumption 3),  
provide  two conditions to avoid arbitrage due to ``manipulator's'' information.
 In the context of a trader and a buyer and with our notation 
 these conditions are:\\
{\em i)} trader's  $Q$ and   buyer's $Q_{bu}$ are mutually absolutely continuous.
 and\\
{\em ii)} for constant manipulator holdings in a short time interval, relative stock prices are a martingale with respect to the information set.  

In this work it has been shown that when the constructed trader's belief probability  is  risk neutral then $Q$ and $Q_{bu}$ are mutually absolutely continuous, i.e., {\em i)} holds.

 

 \subsection{\bf  Binary Statistical Experiments and Calm stock}

\quad By adopting the statistical experiment model  in option pricing, empirical findings are confirmed quantitatively and new information is obtained
for calm stock.

{\em a)} The theory of statistical experiments provides an explanation for volatility's role in the transaction.

Under the assumptions  of Proposition  \ref{t:asydicalm}, from 
 (\ref{eq:traderprob}) 
and (\ref{eq:buyerprob}) 
 it follows 
that  the binary experiment
${\cal E}_{n}=\{P_{tr,n}, P_{bu,n}\}$ 
converges  
to the Gaussian experiment
${\cal G}=\{P_0=N(0,1), \ P_T=N(\sigma_{[t_0,T]},1)\}$
when $n \rightarrow \infty.$ 
From ${\cal G}$'s form  it is
clear that  volatility, $\sigma_{[t_0,T]},$
is  the determining factor in the transaction.

{\em b)} From
 (\ref{eq:traderprob}) and (\ref{eq:buyerprob}),  $Q(v)$ is larger than $Q_{bu}(v)$ for any $v$ and 
 therefore the event $\{S_T>X\}$ has higher probability for the buyer than for the trader. 

{\em c)} Let ${\cal P}^*_{bu}$ be the belief-probability of the buyer corresponding to ${\cal P}^*,$ obtained with the same translations
on $Q_{bu}$, as for ${\cal P}^*$ via $Q.$ 
 The  buyer's price has indeed a volatility premium, with the coefficient $s_{t_0}$ in the  
trader's price's (\ref{eq:Eprice})  replaced by $s_{t_0}e^{\sigma^2_{[t_0,T]}}:$
$$ 
E_{{\cal P}^*_{bu}}e^{-r(T-t_0)}(S_T-X)I(S_T>X)$$
$$
=s_{t_0} e^{\sigma^2_{[t_0,T]}}\Phi(\frac{\ln(s_{t_0}/X)+r(T-t_0)+1.5\sigma^2_{[t_0,T]}}{\sigma_{[t_0,T]}})-Xe^{-r(T-t_0)}\Phi((\frac{\ln(s_{t_0}/X)+r(T-t_0)+.5 \sigma_{[t_0,T]}^2}{\sigma_{[t_0,T]}}).
$$

By reconstructing an agent's  utility function   Ross (2015) recovers the natural probability for the stock price returns,  the pricing kernel and the market risk premium.
Herein probabilities ${\cal P}^*_{bu}$ and  ${\cal P}^*$ are both  normal, with the same variance and the mean of the buyer's 
probability, ${\cal P}^*_{bu},$  is larger than that of ${\cal P}^*.$   The difference between the buyer's price and the price of the option (via Corollary \ref{c:EpriceCalm})  is a measure of the market risk premium.

\section{Appendix 1: Proofs}

\bel \label{p: nocond}  Under assumptions ${\cal A}1-{\cal A}4,$ 
\begin{equation}
\label{eq:nocond}
E_{Q^*}[\exp\{V_{T-t}+ \ln a_{[t,T]}\}|{\cal F}_t]=E_{Q^*}\exp\{V_{T-t}+ \ln a_{[t,T]}\}.
\end{equation}
\enl
 {\bf Proof of Lemma \ref{p: nocond}:} ${\cal F}_t$ is generated  by the countably many stock prices  in $(0,t).$
For every time sequence $0<t_{{m_1^n}}<  t_{{m_2^n}}< \ldots < t_{{m_n^n}}<t,$ that becomes dense in $[0,t]$  as $n$ increases to infinity, 
 the corresponding prices
$$S_{t}, S_{t_{{m_n^n}}},  S_{t_{m_{n-1}^n}}, \ldots, S_{t_{{m_1^n}}}$$
provide the same information as 
$$\frac{S_{t}}{ES_{t}},  \frac{S_{t_{{m_n^n}}}}{E S_{t_{{m_n^n}}}},
\frac{S_{t_{{m_{n-1}^n}}}}{E S_{t_{{m_{n-1}^n}}}},
\ldots,   \frac{S_{t_{{m_1^n}}}}{E S_{t_{{m_1^n}}}}$$
or,
$$\frac{S_{t}/ES_{t}}{ S_{t_{{m_n^n}}}/E S_{t_{{m_n^n}}} },
 \frac{ S_{t_{m_n^n}}/E S_{t_{ {m_n^n}} } }
{S_{t_{{m_{n-1}^n}}}/E S_{t_{{m_{n-1}^n}}}},\ldots, 
\frac { S_{t_{{m_2^n}}}/E S_{t_{{m_2^n}}}}{ S_{t_{{m_1^n}}}/E S_{t_{{m_1^n}}}}, S_{t_{{m_1^n}}}/E S_{t_{{m_1^n}}} $$ 
that coincides with
$$\frac{\frac{S_T/ES_T}{  S_{t_{m_n^n}}/E S_{t_{ {m_n^n}}} }}{ \frac{S_T/ES_T}{S_{t}/ES_{t}}}, 
\frac{  \frac{S_T/ES_T}  {S_{t_{{m_{n-1}^n}}}/E S_{t_{{m_{n-1}^n}}}}}{ \frac{S_T/ES_T} {  S_{t_{m_n^n}}/E S_{t_{ {m_n^n}}} }},\ldots, \frac { \frac{S_T/ES_T} {S_{t_{{m_1^n}}}/E S_{t_{{m_1^n}}}} }{   \frac{S_T/ES_T} 
{S_{t_{{m_2^n}}}/E S_{t_{{m_2^n}}}}}, S_{t_{{m_1^n}}}/E S_{t_{{m_1^n}}}  $$
or, by taking logarithms 
\begin{equation}
\label{eq:unconexpval}
V_{T-t_{m_n^n}} -V_{T-t}, V_{T-m_{n-1}^n}-V_{T-m_{n-2}^n}, \ldots, V_{T-m_{2}^n}-V_{T-m_{1}^n},  S_{t_{{m_1^n}}}/E S_{t_{{m_1^n}}}. 
\end{equation}
When $n$ is large, the last term in (\ref{eq:unconexpval})  approaches unity since both numerator and denominator converge to $s_0$ (from ${\cal A}4$) and  the remaining terms  are increments of a L\'{e}vy process and are therefore independen of $V_{T-t} (=V_{T-t}-V_{T-T}).$ $\hspace{3ex} \hfill \Box$\\
 
{\bf Proof of Proposition \ref {p:QQ*}:} {\em a)} In Proposition \ref{t:LLCmain}  the moment generating function $\psi_{Q_0}(s)$ is determined
for $s \in (0,1).$  The integrand  in 
$\psi_{Q_0}(s)$ (see  (\ref{eq:mgflambda})), 
$$(1+y)^{2s}-1-2sy$$
is bounded by ``some''  multiple of $y^2$ (Le Cam, 1986, p. 465, lines 18-22) and  $E_{L_{tr}}Y^2$ is finite from assumption
 {\cal A}3. From dominated convergence theorem, 
\begin{equation}
\label{eq:limQ0}
\lim_{s \rightarrow 1} \ln  \psi_{Q_0}(s)=\ln \psi_{Q_0}(1)=\ln  E_{Q_0}e^{V_{T-t_0}}=\mu_{[t_0,T]}+.5 \sigma_{[t_0,T]}^2+E_{L_{tr}}Y^2 I(Y\neq 0).
\end{equation}
(\ref{eq:Pln}) holds for $Q_0$ if and only if $ \ln \psi_{Q_0}(1)$ in (\ref{eq:limQ0}) vanishes.

{\em b)} Follows from {\em a)}.

{\em c)} Since {\em i)} the L\'{e}vy measure  $L_{tr}$  has no mass at -1 (by ${\cal A}1$)  and
{\em ii)} Proposition \ref{t:LLCmain} shows $\Delta$-convergence of (a subsequence of)  the experiments
 ${\cal E}_n=\{ P_{tr,n}, P_{bu,n} \}$  to the experiment $\{Q_0, Q_{bu}\},$  it follows that 
$\{ P_{tr,n} \}$ and $\{ P_{bu,n} \}$  are contiguous, i.e. $Q_0$ and $ Q_{bu}$ are mutually absolutely continuous, 
 if and only if, 
\begin{equation}
\label{eq:contiguity}
\lim_{s \rightarrow 1} \psi_{Q_0}(s)=\psi_{Q_0}(1)=1,
\end{equation}
which from (\ref{eq:limQ0})  holds,  if and only if, (\ref{eq:neutraliff}) holds.

d) Due to contiguity,  (\ref{eq:QB2}) holds (see, e.g,  Le Cam and Yang, 1990, p. 22, the Proposition). $\hfill\Box$\\


 {\bf Proof of Proposition \ref{p:STpstar}:} It is enough to show that ${\cal A}1-{\cal A}4$  hold for $t_0 \le t \le T.$  When $S_{t_0}=s_{t_0}$ assumptions ${\cal A}1, {\cal A}2, {\cal A}4$  still hold. For  ${\cal A}3$ observe that since $S_{t_0}=s_{t_0},$
\begin{equation}
\label{eq:ggr1}
0\le E_{P_{t_0}}Y_{n,1}^2=\int (\sqrt{p_{t_1^n}}-1)^2dP=2(1-\int \sqrt{p_{t_1^n}}dP) \le 2.
\end{equation}
From (\ref{eq:ggr1}) it follows that ${\cal A}3(ii)$ holds.

For ${\cal A}3(i)$ to hold  it is enough to show that
$$
\lim_{n \rightarrow \infty} E_{P_{t_0}}Y_{n,1}^2=0
$$
or from (\ref{eq:ggr1}) that
\begin{equation}
\label{eq:ggr3}
\lim_{n \rightarrow \infty} E_P|\sqrt{p_{t_1^n}}-1|=0.
\end{equation}
 Since 
$$ E_P|\sqrt{p_{t_1^n}}-1|\le  E_P|p_{t_1^n}-1|$$
for (\ref{eq:ggr3}) to hold it is enough that 
\begin{equation}
\label{eq:ggr4}
\lim_{n \rightarrow \infty}  E_P|p_{t_1^n}-1|=0. 
\end{equation}
(\ref{eq:ggr4})  follows from:
\bel
\label{l:beautggr}
(Roussas, 2005, Lemma 3, p. 138, 2014, Lemma 3, p. 109) Assume $X_n \ge 0, \ EX_n<\infty, n\ge1.$ Then
$$\lim_{n \rightarrow \infty}  E|X_n-X|=0 \Longleftrightarrow  plim_{n \rightarrow \infty}X_n= X \mbox{ and} 
\lim_{n \rightarrow \infty} EX_n=EX<\infty.$$
\enl
with $X_n=p_{t_1^n}, \ X=1$ since $E_P p_{t_1^n}=1 \  \forall \  n\ge1$  and  ${\cal A}4$ holds. $\hfill \Box$\\


{\bf Proof of Corollary \ref{t:asydicalm}:}
 Follows from (\ref{eq:Q}) and (\ref{eq:QB1}) 
and Le Cam (1986, p. 470) for L\'{e}vy measures $L_{tr}$ and $L_{bu}$ concentrated at $y=0. \hfill \Box$\\

{\bf Proof of Proposition  \ref{p:A3calmgbm}:} {\em a)} From (\ref{eq:gbm}), with $B_t$ standard Brownian motion, we obtain
$$ES_t=s_0exp\{\mu t\}, \hspace{5ex} p_t=e^{-.5\sigma^2t+ \sigma B_t}.
$$
For $t_{j-1}<t_j,$ since $B_{t_j}-B_{t_{j-1}}$ is independent of $B_{t_{j-1}},$ it holds
$$E_{P_{t_{j-1}}}(\sqrt{\frac{p_{t_j}}{p_{t_{j-1}}}}-1)^2 =E (e^{-.25 \sigma^2(t_j-t_{j-1})+.5  \sigma (B_{t_j}-B_{t_{j-1}})}-1)^2
e^{-.5\sigma^2 t_{j-1}+\sigma B_{t_{j-1}}}$$
$$=E (e^{-.25 \sigma^2(t_j-t_{j-1})+.5  \sigma (B_{t_j}-B_{t_{j-1}})}-1)^2 \cdot Ee^{-.5\sigma^2 t_{j-1}+\sigma B_{t_{j-1}}}
=E (e^{-.25 \sigma^2(t_j-t_{j-1})+.5  \sigma    B_{t_j-t_{j-1}} }-1)^2 $$
$$=2(1-e^{-.125\sigma^2(t_j-t_{j-1})}) \sim .25 \sigma^2(t_j-t_{j-1})[1+C  \sigma^2 (t_j-t_{j-1})]$$
for small $t_j-t_{j-1}$
 values; $C$ is a generic bounded constant, bounding the second derivative of  $e^{-.125 \sigma^2(t_j-t_{j-1})}$  in its Taylor expansion
around zero.


It follows that for any partition $t_1, \ldots, t_{k_n-1}$ of $[t_0,T]$ with mesh size $\delta_n$
$$ \sup \{E_{P_{t_{j-1}}}(\sqrt{\frac{p_{t_j}}{p_{t_{j-1}}}}-1)^2 , j=1,\ldots, k_n \} \le .25 \sigma^2 \delta_n[1+ \max\{C_j, \ j=1,\ldots, k_n\} \delta_n] $$
thus, the left side converges to zero when $\delta_n$ converges to zero and ${\cal A}3 (i)$ holds.
Note that $\max\{C_j, \ j=1,\ldots, k_n\} $ is determined by $\delta_n.$

For ${\cal A}3 (ii)$ observe that  for small $\delta_n$
$$|\sum_{j=1}^{k_n}[E_{P_{t_{j-1}}}(\sqrt{\frac{p_{t_j}}{p_{t_{j-1}}}}-1)^2-.25 \sigma^2(t_j-t_{j-1})]|=.25 \sigma^2 \sum_{j=1}^{k_n}C_j(t_j-t_{j-1})^2$$
$$\le .25 \sigma^2 \delta_n \max\{C_j, j=1,\ldots, k_n\}(T-t_0) $$
that  converges to zero when $\delta_n$ converges to zero. 
Thus,
$$\lim_{\delta_n\rightarrow 0} \sum_{j=1}^{k_n} E_{P_{t_{j-1}}}(\sqrt{\frac{p_{t_j}}{p_{t_{j-1}}}}-1)^2=.25 \sigma^2 (T-t_0)$$
 and ${\cal A}3 (ii)$ holds.

{\em b)} From part {\em a)}, a risk neutral probability ${\cal P}^*$ can be obtained via Propositions \ref{t:LLCmain}    and   
\ref {p:QQ*}.  Since ${\cal P}^*$
is unique and $Q$ coincides with ${\cal P}$ as confirmed  in the last sentence of Example  \ref{ex:lognormBSM},
 it follows that the sequences 
$\{{\cal P}_{tr,n}\}$ and $\{{\cal P}_{bu,n}\}$ are contiguous and since for each $n,$ ${\cal P}_{tr,n}$ and ${\cal P}_{bu,n}$ are mutually absolutely continuous, it also  follows that for  $\sum_{j=1}^{k_n} Y_{n,j}$ the variance of the limit is equal to the limit of the variances.
From Le Cam (1986, p. 470, lines -10 to -12), the sums $\sum_{j=1}^{k_n} Y_{n,j}$  have limiting nomal distribution  and the variance of the limit is equal to the limit of the variances if and only if $\sum_{j=1}^{k_n}EY_{n,j}^2 $ converges to a limit and   (\ref{eq:ecalmexp}) holds for every $\epsilon >0.$   
$\hfill \Box$\\

{\bf Proof of Proposition \ref{p:priLevy}:} The option's price, 
\begin{equation}
\label{eq:generalfairprice}
E_{{\cal P}^*}e^{-r(T-t_0)}(S_T-X)I(S_T>X),
 \end{equation}
is  obtained via $Q$ and Propositions \ref{p:QQ*}, \ref{p:STpstar}.

We calculate separately  each of  the two expected values  in (\ref{eq:generalfairprice}) excluding constants.
$$E_{{\cal P}^*} I(S_T>X)
= P^*(\ln \frac{S_T}{s_{t_0}} > \ln \frac{X}{s_{t_0}}) $$
$$=1-Q^*(\ln \frac{X}{s_{t_0}} - \ln a_{[t_0,T]})=1-Q(\ln \frac{X}{s_{t_0}}-r(T-t_0))$$
\begin{equation}
\label{eq:Rtr}
=1-\int \Phi(\frac{\ln \frac{X}{s_{t_0}}-r(T-t_0)+.5\sigma_{[t_0,T]}^2+ E_{L_{tr}}Y^2I(Y\neq 0)-y}{\sigma_{[t_0,T]}})L_{tr,Pois}(dy),
\end{equation}
with the penultimate and the last  equalities obtained using, respectively, (\ref {eq:Q*}) and (\ref{eq:Q}).
$$ E_{{\cal P}^*}S_T I(S_T>X)=s_{t_0}E_{Q^*}e^{V_{T-t_0}+\ln a_{[t_0,T]}}I(V_{T-t_0}>\ln \frac{X}{s_{t_0}} - \ln a_{[t_0,T]})$$
$$=s_{t_0}\int_{\{v>\ln(X/s_{t_0}) - \ln a_{[t_0,T]} \}}  e^{v+\ln a_{[t_0,T]}} dQ(v+\ln a_{[t_0,T]}-r(T-t_0))$$
$$=s_{t_0}e^{r(T-t_0)}\int_{\{w>\ln(X/s_{t_0})-r(T-t_0) \}}  e^w dQ(w)=s_{t_0}e^{r(T-t_0)}\int_{\{w>\ln(X/s_{t_0})-r(T-t_0) \}}dQ_{bu}(w)$$
$$
=s_{t_0}e^{r(T-t_0)}[1-Q_{bu}(\ln(X/s_{t_0})-r(T-t_0))] 
$$
\begin{equation}
\label{eq:Rbu}
=s_{t_0}e^{r(T-t_0)} [1-\int \Phi(\frac{\ln(X/s_{t_0})-r(T-t_0)-.5\sigma_{[t_0,T]}^2- E_{L_{tr}}Y^2I(Y \neq 0)-y}{\sigma_{[t_0,T]}})L_{bu,Pois}(dy)],
\end{equation}
with the second, the penultimate and the last  equalities obtained using, respectively, (\ref {eq:Q*}) ,  (\ref {eq:QB2})  and
 (\ref {eq:QB1}).

Replacing  (\ref{eq:Rtr}), (\ref{eq:Rbu}) in (\ref{eq:generalfairprice}), ${\cal P}^*$-price (\ref{eq:Eprice}) is obtained. $ \hfill \Box$

\section{Appendix 2:  Conditions for Assumption ${\cal A} 3 $ to hold}

\quad Quadratic mean differentiability conditions are sufficient
for ${\cal A} 3 $ to hold. Quadratic mean differentiability holds frequently
in parametric statistical models, e.g., for the normal and log-normal models;
see  Le Cam (1970) and Roussas (1972, Chapter 2).
Let $(\Omega, {\cal F}, {\cal P})$ be a probability space
and let $\rho (t), t \in [0,T],$
be a process indexed by $t.$

\bef The process
$\rho$ is differentiable at $t$ in ${\cal P}$-quadratic mean if
there is $U_t,$ its derivative at $t,$ such that
\begin{equation}
{\frac{1}{\delta^2}\int[\rho(t+\delta)-\rho(t)-\delta U_{t}]^2
d{\cal P}}_{\delta \rightarrow 0} \longrightarrow 0.
\label{eq:defqmd}
\end{equation}

When $t=0$ (resp. $T$) the limit in (\ref{eq:defqmd})
is taken for $\delta$ positive (resp. negative).
\enf

For the prices-densities $\{p_t, \ t \in [0,T]\},$ let
\begin{equation}
\xi(t)=\sqrt{p_t}, \ t \in [0,T].
\label{eq:xi}
\end{equation}

\bep \label{p:forA3tohold} Assume that
$\xi(t)$ is
${\cal P}$-quadratic mean differentiable in
$[t_0,T]$ with derivative $U_{t},$ and that $\sup_{t \in [t_0,T]}
E_{{\cal P}} U_{t}^2 <\infty.$
Then, ${\bf {\cal A}}3 $ holds for the price-densities $p_{t}, \
t \in [0,T].$
\enp


{\bf Proof:} For any $t$ in $[t_0,T]$ and $\delta$ small,
$$\int [\xi(t+\delta)-
\xi(t)]^2d{\cal P}
\le 2[\int [\xi(t+\delta)-\xi(t)-\delta U_{t}]^2d{\cal P}
+ \delta^2
\int U_{t}^2d{\cal P}]$$
$$=2 \delta^2[E_{{\cal P}} U_{t}^2+o(1)].$$
Uniform boundedness  
of $E_{{\cal P}}U_{t}^2$ implies ${\cal A}3(i)$ holds.

For transaction times with small mesh size in $[t_0, T],$
$$\sum_{j=1}^{k_n} \int [\xi(t_i)-
\xi(t_{i-1})]^2d{\cal P}
\le 2 [\sum_{j=1}^{k_n}(t_j^n-t^n_{j-1})^2
E_{{\cal P}}U_{t^n_{j-1}}^2+
o(1)\sum_{j=1}^{k_n}(t_j^n-t^n_{j-1})^2]$$
$$\le 2 (T-t_0)^2 [\sup_{t \in [t_0,T]}
E_{{\cal P}} U_{t}^2 + o(1)] <\infty.$$
$\hspace{3ex} \hfill \Box$

\end{document}